\documentclass[aps,pra,showpacs,twoside,twocolumn,10pt,floatfix]{revtex4-2}
\usepackage[colorlinks=true, citecolor=red, urlcolor=blue ]{hyperref}
\usepackage{times,epsfig,amssymb,amsfonts,amsmath,bm,subfigure,mathtools,amsthm,braket,soul,enumitem,xcolor, physics, graphics,graphicx}
\usepackage[normalem]{ulem}
\usepackage{comment, appendix}
\usepackage{epstopdf}

\usepackage{comment}

\begin{document}

\title{Creating two-qudit maximally entangled quantum link through bulk}

\author{Keshav Das Agarwal$^{1,2}$, Sudip Kumar Haldar$^{1,2,3}$, Aditi Sen(De)$^{1,2}$}
\affiliation{$^1$Harish-Chandra Research Institute, Chhatnag Road, Jhunsi, Prayagraj - 211019, India}
\affiliation{Homi Bhabha National Institute, Training School Complex, Anushakti Nagar, Mumbai - 400094, India}
\affiliation{$^2$Department of Physics and Material Science \& Engineering, Jaypee Institute of Information Technology, Noida - 201304, India}


\begin{abstract}

We design a setup for creating almost maximally entangled two-qudit link between distant nodes which are weakly coupled with the interacting spin-\(s\) bulk (processor). We exhibit that such a quantum link between spins of arbitrary spin quantum number can be formed when the system is prepared at sufficiently low temperature. We find that the Heisenberg and the bilinear-biquadratic (BBQ) spin-\(1\) models are the potential candidates to obtain long-distance near-maximal entanglement in equilibrium. Beyond the static regime,  we further show that initializing the bulk in a fully polarized state and the link in an appropriate qudit state allows the system to dynamically evolve into a highly entangled state under both Heisenberg and BBQ Hamiltonians. Our findings reveal that both the static and the dynamical protocols presented here remain efficient towards creating highly entangled two-qudit states even if the spin quantum numbers of the bulk is lower than that of the link.

\end{abstract}

\maketitle

\section{Introduction}

\begin{figure}
    \centering
    \includegraphics[width=\linewidth]{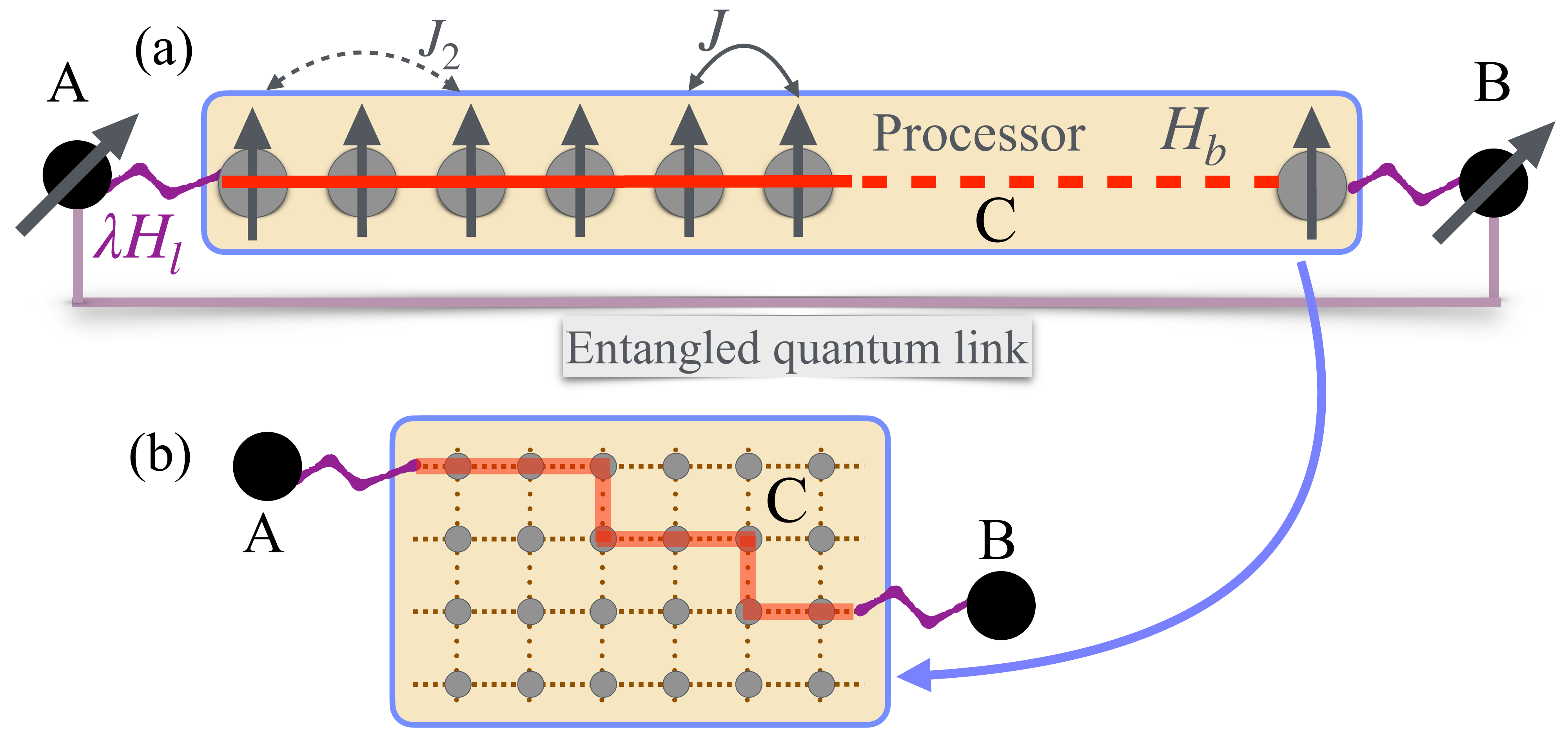}
    \caption{Schematic diagram of a generic quantum processor attached to two distant sites to be entangled, referred to as the entangled quantum link. Quantum processor (bulk) $C$ consists of (a) interacting spins $s_b$ in a chain which can be a part of (b) two-dimensional lattice with controllable interactions. Two distant spin-$s_\ell$ systems at the link sites, denoted as \(A\) and \(B\), are weakly coupled to the bulk. The goal is to obtain a highly entangled quantum link between \(A\) and \(B\), both at zero and nonzero temperatures and through unitary evolution.} 
    \label{fig:spin-chainschematic}
\end{figure}

Entanglement \cite{HoroRMP}, a purely quantum concept~\cite{Schrodinger_1935}, is a central resource for quantum technologies like communication and computation~\cite{nielsenbook}. Entangled photons enable long-distance communication via optical fibers~\cite{RMP12Zukowski}, yet transmitting across distant qubits remains challenging~\cite{Cirac_PRA_1999, repeater_PRL_1998, Repeater_PRA_1999}. Short-range communication is crucial for connecting processors into modular scalable quantum computers~\cite{Kielpinski2002, Blinov2004, Skinner_PRL_2003}.
In this context,  spin chains emerge as an acceptable alternative to photons for obtaining long-distance entanglement (LDE) between two sites~\cite{Bose,venuti_prl}.
Specifically, it was demonstrated that by tuning system parameters appropriately, two spin-\(1/2\) systems, called \(A\) and \(B\), separated by some distance, can share a near-maximally entangled state when they  interact with an interacting spin chain, referred to as the bulk, 
creating a resourceful quantum link at low temperature (see Fig. \ref{fig:spin-chainschematic})~\cite{Wojcik05, venuti_prl, VenutiPRA07, Giampaolo_2010, Lukin11, apollaro_pra_2012, Dhar_2016}. Note that the monogamy of entanglement \cite{wooters_monogamy_pra_2000} demands that the interaction between \(A\) (and \(B\)) and the edge spins of the bulk,
be weaker than the typical interaction between the spins contained in the bulk~\cite{venuti_prl}. 

Prominent examples that possess LDE include antiferromagnetic spin systems, such as the Heisenberg and the \(XXZ\) spin chain with alternating interactions ~\cite{venuti_prl, HU2022}, variations of the Heisenberg spin-$1/2$ ~\cite{Ferreira_PRA_2008} and spin-$1$ chain~\cite{venuti_PRL_2005} with bilinear-biquadratic interaction~\cite{venuti_prl, bbq_prb_2022} and the Fermi-Hubbard model~\cite{Abaach2023}. 
The presence of LDE in such models provide a method to produce entangled link with the same spin-dimension as the bulk in the static situation. However, energy gaps above the ground state in such systems vanish exponentially with the increase of the length of the spin chain~\cite{venuti_prl}. This suggests that LDE can only be found, and be used to create an entangled link at absolute zero temperature, which is physically untenable. Alternatively, the ground state of certain  spin chains like the open \(XX\) model with weak end bond interactions can have a kind of long-distance entanglement, referred to as ``quasi-long-distance" entanglement, between the end spins which decay very slowly with the length of the chain~\cite{Dhar_2016}.
Experimentally, the entanglement between two spins approximately $220\!-\!250 \text{\AA}$ apart has been created through a spin chain prepared from Sr$_{14}$Cu$_{24}$O$_{41}$ block materials below $2.1$K in laboratories~\cite{Nature-expt}. Similarly, 
a nitrogen-vacancy (NV) center-based link connected by a quantum channel was proposed \cite{Lukin11} (see also~\cite{Yao2012}). Most prior works, however, primarily focused on preparing two-qubit maximally or highly entangled quantum connection in equilibrium. 
In contrast, our study goes beyond both equilibrium setups and dimensional restrictions, thereby broadening the scope of entanglement generation and analysis.

Higher-dimensional quantum systems  provide enhanced resources~\cite{gisinBell, cjlmp_2002, son_lee_kim_prl2006}, making them valuable for diverse quantum information tasks, including quantum cryptography \cite{Walborn_prl_2006,cerf_qkd_prl_2002, QKDqudit2, quditQtech},  quantum computation \cite{QCqudits,Joo2019,Wang2020,Nikolaeva2024} and quantum thermal devices \cite{wang2015,Dou2020, santos2019, usui2021, ghosh2022, konar2023}. Despite additional challenges  in controlling qudits  (cf. \cite{atoms_prl_2013, Kues2017, Chi2022}), they can potentially reduce the need for multiple qubit-based resources, offering greater efficiency and flexibility. Moreover, the experimental quantum platforms are naturally multilevel quantum systems, and qubits are typically realized by restricting control to two of the most accessible energy levels.  
Further, dimensional advantages in quantum protocols have been experimentally demonstrated by using several physical substrates including photons \cite{joo_2007,nagali2010, Bouchard2017, Bouchard2018, Imany2019, photonexp}, trapped ions\cite{qutrit_ion_2003,Low2020},  NV center \cite{NVcentrequdits}, superconducting circuits \cite{supercondqudits} and  even synthetic dimensions enabling many-body simulations on single atoms \cite{Ozawa2019, Kanungo2022}.

In this work, we begin by asking the following question: {\it ``Is it possible to create a two-qudit entangled link using a bulk made of multiple interacting spins?"} This question is especially relevant if quantum processors are constructed with qudit or qubit states. We answer affirmatively by demonstrating that {\it both the LDE at low but finite temperatures, as well as the dynamical evolution, can generate entangled links through suitable tuning of the system parameters}.  This raises an immediate question: {\it Is it necessary to have the bulk and quantum link composed of spins with the same spin quantum number to achieve entanglement?} We determine that it is not necessary for establishing a highly entangled quantum link. In particular, we report that a near-maximally entangled two-qudit state emerges at low temperatures when a spin-\(1/2\) Heisenberg interacting bulk is weakly coupled to a  spin-\(s_\ell\) link spins. However, when the bulk is replaced by spin-\(1\)  bilinear-biquadratic Hamiltonian~\cite{Lai1974,sutherland1975, Babujian1982, fath1991, fath1993, bbq_prb_2022} and the link spins follows the same weak interactions with the bulk, the near-maximal entanglement can be obtained with greater control of the system compared with the spin-\(1/2\) bulk system by suitably adjusting the system parameters. Up to a moderate system sizes, all the results reported here remain independent of system size.

Beyond equilibrium, we demonstrate that a highly entangled quantum link between two distant processors can be dynamically generated through unitary evolution from initially fully separable states. The advantage of this dynamical approach is that the  unitary gates on the qudits have already been experimentally realized \cite{gao_pra_2019,Imany2019,Low2020,Chi2022,Hrmo2023,liu_prx_2023,fischer_PRXQuantum_2023}, thus eliminating the need for the temperature control required in the static scenario, provided that decoherence effects remain minimal. Specifically, we prepare the link spins in arbitrary separable states and the bulk spins in a polarized product state, ensuring that the entire system is fully separable, and then evolve it under the same Hamiltonian used in the static case. In this situation, we identify the optimal conditions for achieving a highly entangled link by optimizing over both the time and initial state parameters.

The organization of this paper is as follows. In Sec.~\ref{sec:framework}, we describe the framework for creating an entangled quantum link in both equilibrium and nonequilibrium scenarios. The framework consists of a description of the model Hamiltonian, the definition of the entanglement measure - logarithmic negativity \cite{vidal_pra_2002}, and a discussion of the numerical techniques used in this work. Section~\ref{sec:equi} discusses our findings on achieving near-maximally two-qudit quantum entangled link or long-distance entanglement in an equilibrium scenario through a bulk of spin-$s_b$ chain (\(s_b =1/2, 1\)). In Sec.~\ref{sec:dynamics}, we report how to dynamically create a highly entangled quantum link between two distant nodes  in the same setup. The results are summarized in Sec. \ref{sec:conclu}.

\section{Framework of Building Entangled Links}
\label{sec:framework}

The quantum processors are generically created as a two-dimensional array of interacting spins with controllable interaction strengths. The generic setup for establishing the entangled quantum link between two distant spins, with such processors, can be conceived as follows. Consider  a bulk (processor) $C$ consisting of interacting spins (with spin quantum number \(s_b\)) in a one-dimensional array, which is generated by considering its decoupling from the rest of the spins of \(C\). The two-distant spin-$s_\ell$ systems (link spins), denoted as \(A\) and \(B\), are weakly coupled to the two opposite ends of $C$, as illustrated in Fig.~\ref{fig:spin-chainschematic}. Since different arrays of spins in the processor are effectively decoupled from the rest of the spins of the lattice and the link, one may approximately consider a one-dimensional  \(N-2\) spin-\(s_b\) system as the bulk as shown in Fig.~\ref{fig:spin-chainschematic}(a), which we will do in this work. Let the density matrix of the entire system be \(\rho_{ABC}\) of \(N\) parties, and we are interested in investigating \(\rho_{AB}\) after tracing out the bulk \(C\). We aim to find the optimal equilibrium and dynamical conditions for  maximum entanglement between \(A\) and \(B\) by tuning the interactions between the bulk spins and the link spins of arbitrary spin quantum number. In a static situation, LDE has been obtained in the literature when either \(s_b = s_\ell=1/2\) or \(s_b = s_\ell = 1\) is at zero temperatures, i.e., with the ground states~\cite{Wojcik05, venuti_prl, VenutiPRA07, Giampaolo_2010, Lukin11, apollaro_pra_2012, Dhar_2016}, or at low temperatures~\cite{Ferreira_PRA_2008}. 
Here, we are interested in finding the interaction strength and other system parameters necessary to obtain the near-maximal entanglement of \(\rho_{AB}\) even when $s_\ell \neq s_b$, both in equilibrium and nonequilibrium situations. Note here that the Hamiltonian, which has LDE for \(s_b = s_\ell=1/2\), may not necessarily provide near-maximal entanglement for $s_\ell \neq s_b$ at low temperature. It is indeed the case, as we find for the case spin-$1$ and spin-$3/2$ $XY$ link spins interacting with the spin-$1/2$ $XX$ Hamiltonian in the bulk~\cite{Dhar_2016}, highly entangled quantum links cannot be obtained (see Appendix~\ref{app:XY}). 

\subsection{Model Hamiltonian: Bulk and link} \label{subsec:Hamiltonian}
 
Let us now describe the Hamiltonian for the entire setup. The bulk \(C\), an effective one-dimensional chain of spin-\(s_b\), interacts via the bilinear-biquadratic (BBQ) interaction, given by 
\begin{eqnarray}
    H_{i,j} =& \begin{cases}
            \vec{S}_i\cdot\vec{S}_{j} \quad\quad\;\text{ if } i^{\text{th}} \text{ or } j^{\text{th}} \text{ site is spin-}1/2  \\
            \cos\theta\vec{S}_i\cdot\vec{S}_{j}+ \sin\theta(\vec{S}_i\cdot\vec{S}_{j})^2 \:\text{ otherwise}
            \end{cases}\hspace{-4mm},
            \label{eq:bilinbiq}
\end{eqnarray}
where $\vec{S}_i=\hat{S}^x_i \hat{x} + \hat{S}^y_i \hat{y} + \hat{S}^z_i \hat{z}$ is a quantum spin operator in a three-dimensional space acting at the site $i$. Note that  it is SU(2) invariant.
The link spins \(A\) and \(B\) are two single spin-\(s_\ell\) systems which are weakly connected to two opposite ends of \(C\) through the BBQ interaction (as shown in Fig. \ref{fig:spin-chainschematic}).
The Hamiltonian of the bulk and edge spins together reads as 
\begin{align}
    H^{(s_\ell,s_b)}(\lambda) &= J H^{s_b}_{b} + \tilde{\lambda} H^{(s_\ell,s_b)}_{l},  \text{with}\\
    H^{s_b}_{b} &= \sum_{i=2}^{N-2} H_{i, i+1}+ J_2\sum_{i=2}^{N-3} H_{i, i+2}, \label{eq:Hdef2} \\
    H^{(s_\ell,s_b)}_{\ell} =& H_{1,2} + H_{N-1, N} + J_2(H_{1,3} + H_{N-2,N}), 
            \label{eq:Hdef4}
\end{align}
where  the dimensionless parameter $\lambda=\frac{\tilde{\lambda}}{J}$ gives the relative coupling strength between \(A\) (or \(B\)) and the bulk, and $J_2$ is the strength of the next-nearest neighbor (NNN) interaction. Note further that when either of the sites $i$ or $j$ is a spin-$1/2$ system,  $(\vec{S}_i\cdot\vec{S}_{j})^2 = \frac{1}{2} (\mathbb{I}-\vec{S}_i\cdot\vec{S}_{j})$. Therefore, in such cases, there is no biquadratic term and the interaction reduces to the Heisenberg interaction.  Similarly, for spin $s_\ell=\frac{1}{2}$, the coupling between \(A\) (or \(B\)) and \(C\) is also taken to be Heisenberg.  Here, we study both the thermal equilibrium state of the Hamiltonian \( H^{(s_\ell,s_b)}(\lambda)\), and the dynamical state, obtained from its unitary time evolution.

\subsection{Quantifying entanglement between two link spins} \label{subsec:ent}

To ensure that the state \(\rho_{AB}\) of the link spins (obtained after tracing out \(C\)), possesses almost maximal entanglement, we use logarithmic negativity (LN) \cite{vidal_pra_2002, plenio_prl} as an entanglement measure. 
For the bipartite state $\rho_{AB}$, it is defined as 
\begin{eqnarray}
    \tilde{\mathcal{E}} = \log_2[2\mathcal{N}(\rho_{AB})+1]. \nonumber
\end{eqnarray}
Here the negativity, $\mathcal{N}(\rho_{AB})$  is the absolute sum of the negative eigenvalues of the partially transposed state $\rho_{AB}^{T_A}$ (or $\rho_{AB}^{T_B}$) with the transpose being taken either over the party \(A\) (or \(B\)) \cite{peres_prl_1996, horodecki_pla_1996}. Although $\tilde{\mathcal{E}}=0$ is only a necessary condition of separability for $d^2_\ell>6$ with $d_\ell (= 2 s_\ell +1)$ being the dimension of the Hilbert space of individual link spins $s_\ell$ at \(A\) and \(B\) \cite{PPTboundPRL, NPTbound00, NPTboundbruss}, it is also a sufficient condition for a large class of SU(2) symmetric states \cite{Breuer_2005}. Nonetheless, a nonvanishing $\tilde{\mathcal{E}}$ confirms that the state $\rho_{AB}$ is entangled. Further, LN is maximum ($=\log_{2}d_\ell$) for $\rho_{AB}=|\Phi^{+}_{AB}\rangle\langle \Phi^{+}_{AB}|$ ($|\Phi^{+}_{AB}\rangle = \frac{1}{\sqrt{d_{\ell}}}\sum_{k=0}^{d_\ell-1}|kk\rangle$) and its local unitarily equivalent states with $\{|0\rangle, |1\rangle,\ldots |2s\rangle\}$ representing the $S^z$ eigenbasis of spin-$s$, with $|0\rangle$ corresponding to the maximum eigenvalue. For a fair comparison between different $s_{\ell}$, we scale $\tilde{\mathcal{E}}$ of the bipartite state $\rho_{AB}$ of the link sites having individual dimension $d_\ell$ as $\mathcal{E} = \frac{1}{\log_{2}d_\ell}\tilde{\mathcal{E}}$ such that \(0 \leq \mathcal{E} \leq 1\) for all $d_{\ell}$.

\subsection{Numerical simulation technique} \label{subsec:num_sim}

The spin models, both in equilibrium and nonequilibrium situations, considered in this work cannot be solved analytically since we will be dealing with higher-dimensional systems and the Hamiltonian involves next-nearest neighbor (NNN) interactions along with the nearest-neighbor (NN) ones. 
Let us briefly discuss the numerical methods that we carry out to perform the entire analysis.

{\it Equilibrium scenario.} We use the exact diagonalization procedure based on the Lanczos method \cite{Lanczos, arma2016, arma2019} to find the low-lying states of the model, which are closely spaced in our case ($\lambda \ll 1$). In the extreme case $\lambda= 0$, $H(\lambda)$ [$\equiv H^{(s_\ell,s_b)}(\lambda)$] has the ground-state degeneracy of $\sim d_\ell^2$. Therefore, when $\lambda\to0$, $H(\lambda)$ has $\sim d_\ell^2$ low-lying eigenstates.
To obtain the $N$-party thermal state, $\rho_{ABC} = e^{-\beta H(\lambda)}/\text{tr}(e^{-\beta H(\lambda)})$, with \(\beta = 1/K_BT\) being the inverse temperature, \(K_B\) being the Boltzmann constant, we compute $M+1$ low-lying eigenstates $\{\ket{E_j}_{ABC}\}_{j=0}^{M}$ with eigenvalues $\{E_j\}$ of $H(\lambda)$ in the ascending order, giving $E_0$ as the ground-state energy. Since we consider low temperatures, i.e., high $\beta\geq10^3$, we obtain $\rho_{ABC}\approx\sum_{j=0}^{M} e^{-\beta E_j} |E_j\rangle \langle E_j|/Z$, with $Z=\sum_{j=0}^{M} e^{-\beta E_j}$ while choosing $M$ (with $M+1\geq 3d_\ell^2$) such that $e^{-\beta(E_M-E_0)}<\epsilon\sim O(10^{-6})$.

We also compute the purity of the thermal state as $\mathcal{P}=\frac{1}{Z}\sum_{j=0}^{M}\left(e^{-\beta E_j}\right)^2 \in (0,1]$.
When $\beta(E_1-E_0)>>1$, $\mathcal{P}\sim1$ indicating that the unique ground state is obtained at a sufficiently low temperature such that the energy gap between the ground state and the first excited state is much larger than the temperature in the energy scale.

{\it Nonequilibrium scenario.} In this dynamical scenario, we look at the entanglement generation capability of the model.
Initially, the system is prepared in a fully separable state, \(|\Psi_0\rangle_{ABC} \equiv |\Psi (t=0)\rangle_{ABC} = \otimes_{j=1}^N |\phi_j\rangle \). Then the unitary evolution occurs according to the Hamiltonian \(H(\lambda)\), leading to the  
evolved state $|\Psi(\lambda, t)\rangle_{ABC}=e^{-i H(\lambda) t}|\Psi_0\rangle_{ABC}$.
Here, we assume the system to be effectively isolated, free from environmental disturbances and decoherence, so that its dynamical evolution remains purely unitary.
In this situation, we compute the reduced density matrix $\rho_{AB}(\lambda, t)$ from the $N$-party dynamical state $\rho_{ABC}(\lambda, t)=|\Psi(\lambda, t)\rangle_{ABC}\langle\Psi(\lambda, t)|$. 

To handle the situation numerically, we apply the Chebyshev series approximation to the unitary operator \cite{Chb_rmp06, Chb_SciPostPhys19}, where a finite number of terms (at least $30$) are taken, ensuring that the magnitude of the last term is $<10^{-8}$.  Note that it only requires the action of the Hamiltonian on an arbitrary state and  the number of the terms in the series increases linearly with the increase of the Hilbert space dimension and the time step of the evolution and hence this technique is efficient.

\begin{figure}
    \centering
    \includegraphics[width=\linewidth]{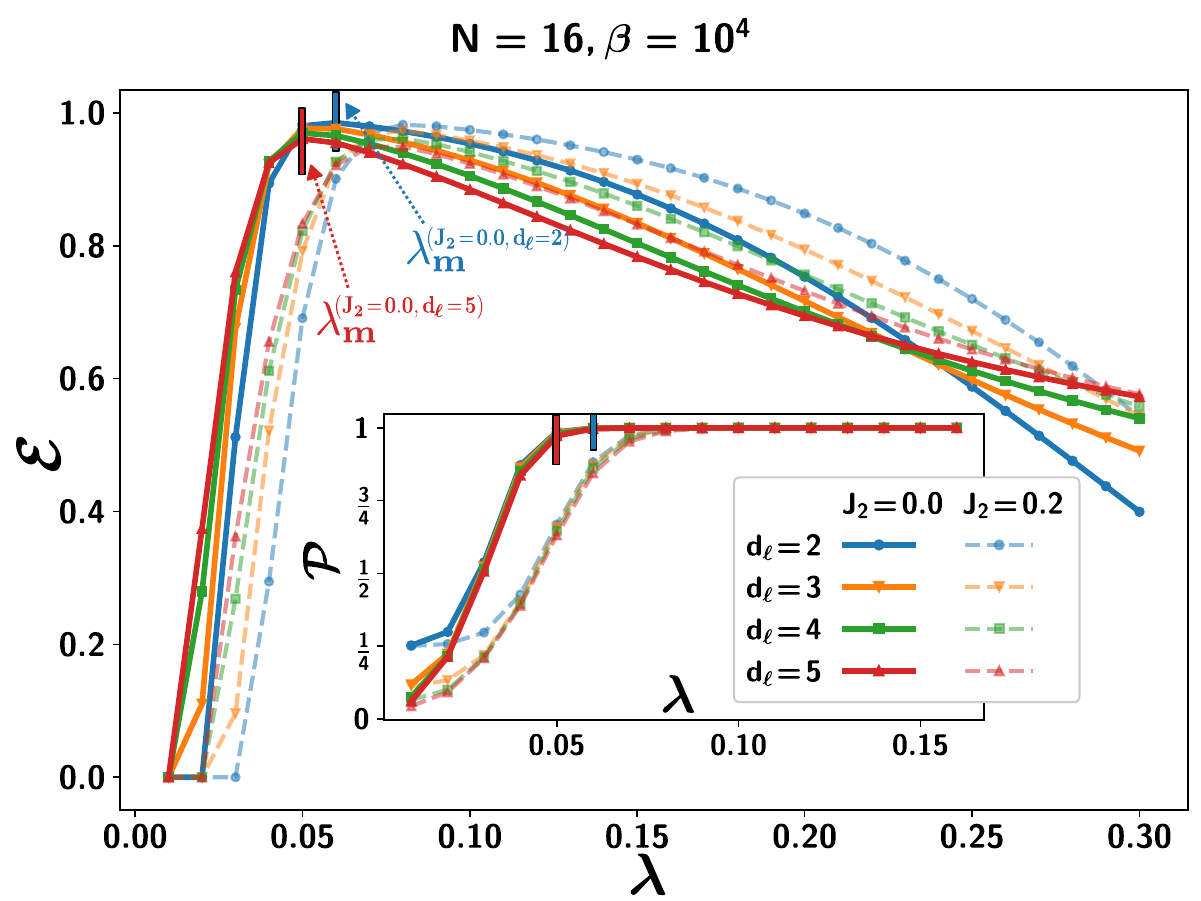}
    \caption{Entanglement $\mathcal{E}$ (ordinate) of the quantum link, against the interaction strength $\lambda$ (abscissa) for different spin dimensions of the link site, \(d_\ell\). The interactions between  spin-$\frac{1}{2}$ bulk sites and bulk-link pair  are governed by the Heisenberg Hamiltonian with and without the next-nearest interaction strength $J_2$. The maximum entanglement, \(\mathcal{E}^{\max}\) of the thermal state, achieved at \(\lambda_m\), decreases slightly with the increase of \(d_\ell\), while  in the presence of $J_2$, \(\lambda_m\) increases.(Inset) Purity $\mathcal{P}$ for the corresponding system with \(\lambda\) having vertical bars at $\lambda_m$. Here $N=16, \beta=10^4$ and all quantities are dimensionless.}
    \label{fig:b_2_ent}
\end{figure}

\begin{figure*}
    \centering
    \includegraphics[width=\textwidth]{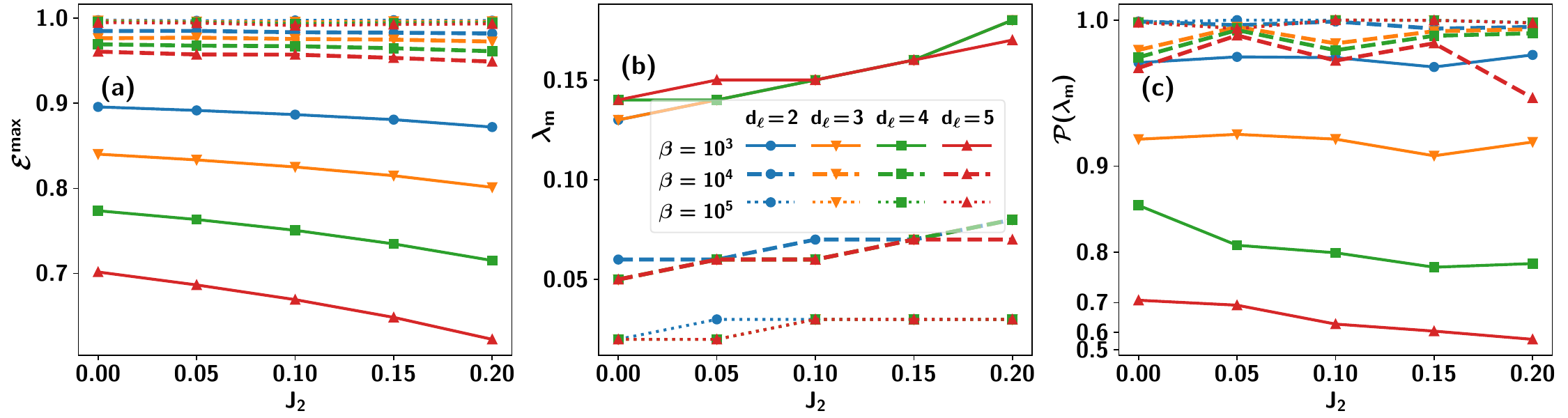}
    \caption{Establishing entangled quantum link with the variation of the NNN interaction strength $J_2$ (abscissa) in spin-$1/2$ bulk ($s_b=1/2$).  (a) $\mathcal{E}^{\max}$ (ordinate) represents the maximum entanglement that can be prepared between distant link spins after optimizing over the interaction strength \(\lambda\), representing interaction between the link spins and the edge of the bulk. (b) Edge interaction strength $\lambda_m$ (ordinate) where $\mathcal{E}^{\max}$ is achieved against \(J_2\). (c) Purity $\mathcal{P} (\lambda_m)$ (ordinate) at the edge interaction strength $\lambda_m$ with \(J_2\).  Different linestyles represent different inverse temperature values, and different markers denote different spin-dimensions of the link spins.
    The system size is $N=16$. All quantities are dimensionless.}
    \label{fig:b_2_ent_max_J2}
\end{figure*}

\begin{figure}
    \centering
    \includegraphics[width=\linewidth]{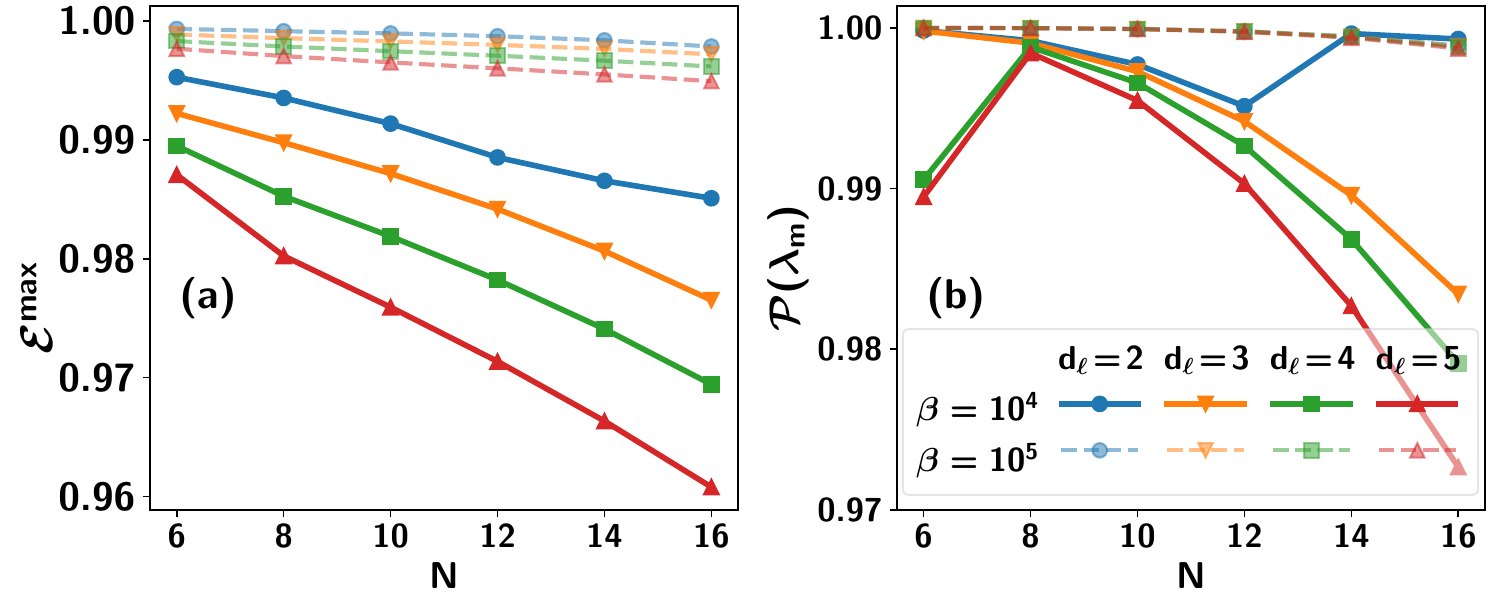}
    \caption{Finite-size effects on entangled link. (a) $\mathcal{E}^{\max}$ (ordinate) against system size $N$ (abscissa)  and (b) purity $\mathcal{P}(\lambda_m)$ at $\lambda_m$ (ordinate) versus \(N\). Here $J_2=0.0$. It is evident that with the increase of \(\beta\), $\mathcal{E}^{\max}$ does not depend on the system size. All other specifications are same as in Fig. \ref{fig:b_2_ent_max_J2}. All quantities are dimensionless.}
    \label{fig:b_2_ent_max_N}
\end{figure}

\section{Two-qudit entangled link Through Bulk Spin-$s_b$ Chain in Equilibrium - Dimensional benefit}
\label{sec:equi}

We now explore the creation of a near-maximally entangled quantum link by adjusting system parameters at near-zero temperatures. In this work, we consider two bulk systems, the (i) spin-$1/2$ and (ii) spin-$1$ chain to examine if there is any dimensional advantage in replacing the qubits with qudits in the bulk.

\subsection{Spin-$\frac{1}{2}$ bulk versus spin-$s_\ell$ quantum link}

Let us first consider the system with spin-$\frac{1}{2}$ bulk \(C\) for various spin-$s_\ell$ link spins at \(A\) and \(B\). It was shown that the near-maximal entanglement between the link spins could be established approximately by tuning the system parameters when the spin-$\frac{1}{2}$ bulk follows only the nearest-neighbor $XX$ interaction while the spin-$\frac{1}{2}$ link spins are connected to the bulk via the nearest-neighbor $XY$ interaction~\cite{Dhar_2016}. We find that such interactions fail to possess LDE for $s_b=1/2$ and $s_{\ell}>1/2$ (see Appendix~\ref{app:XY}). Hence, we consider the Heisenberg interaction between the bulk spins and also between the bulk and the link spin, as also discussed in Sec. \ref{subsec:Hamiltonian}.

We find that the antiferromagnetic Heisenberg interactions between the nearest-neighbor sites (i.e., by fixing $J_2=0$) can almost maximally entangle the link spins \(A\) and \(B\) irrespective of the value of $s_\ell$. However, for a very small coupling, i.e., when $\lambda < \lambda_v$ with $\lambda_v$ being the minimum link interaction strength to have nonvanishing entanglement between the link spins, we notice that at a given $\beta$, the number of closely spaced low-lying states $\propto d_\ell^2$ reduces the purity $\mathcal{P}$ of the thermal state $\rho(\beta, \lambda)$, resulting in a reduction in  entanglement between \(A\) and \(B\). For example, at $\beta = 10^4$, the purity $\mathcal{P}$ is $\frac{1}{4}$ for $d_\ell=2$ and $\lambda=0.01$, which, in turn, possesses a vanishing entanglement between the two-qubit link state (see Fig.~\ref{fig:b_2_ent}). Further, we notice that the $\lambda_v$ decreases with the increase of dimensions $d_\ell$ of the individual site of the link (see Fig.~\ref{fig:b_2_1_ent} of Appendix~\ref{app:subsec:stat} for $\beta=10^3$). As $\lambda$ is increased, initially $\mathcal{P}$ increases very slowly with practically no gain in entanglement although, with moderate values of $\lambda$, $\mathcal{P}$ starts increasing faster with $\lambda$ and $\mathcal{E}$ also starts growing. $\mathcal{E}$ reaches a maximum, denoted as $\mathcal{E}^{\max}=\max_{\lambda}\mathcal{E}$ at $\lambda=\lambda_m$. For various $\beta$ ($\beta=10^3, 10^4$ and $10^5$ in Fig.~\ref{fig:b_2_ent_max_J2}), three immediate observations emerge: (i)  $\mathcal{E}^{\max}$ decreases with the increase of $d_\ell$ for both \(\beta = 10^3\) and \(10^4\) although, the decrement of $\mathcal{E}^{\max}$ is much more prominent for \(\beta=10^3\) compared to the case with \(\beta=10^4\); (ii) the behavior of $\mathcal{E}^{\max}$ is directly connected to the purity of the entire state 
i.e., when $\mathcal{P}(\lambda_m)\sim1$, $\mathcal{E}^{\max}$ is obtained;  (iii) $\lambda_m$ almost remains constant with $d_\ell$ as also shown in Fig.~\ref{fig:b_2_ent_max_J2}. All the observations depend on the temperature chosen. Specifically, very low temperatures can erase any differences seen due to the increase of individual dimensions of the link node (see Fig.~\ref{fig:b_2_1_ent} of Appendix~\ref{app:subsec:stat}). We also notice that $\mathcal{E}$ decreases further with the increase of $\lambda$, 
although $\mathcal{P}$ still continues to increase slowly. This is because 
the monogamy property of entanglement \cite{wooters_monogamy_pra_2000} prevents a large entanglement between the two sites at \(A\) and \(B\) in the presence of strong coupling with spins $s_b$ of  \(C\). Therefore, there is a trade-off between coupling strength $\lambda$ and the inverse temperature $\beta$ to achieve high entanglement $\mathcal{E}^{\max}$ between the two link spins. At $\beta=10^4$, $\mathcal{E}^{\max}\sim 1$ is achieved for all link spins (Fig. \ref{fig:b_2_ent}), with $\mathcal{P} (\lambda_m)\sim 1$ even at small $\lambda$ ($\lambda\sim0.05$). It implies that a unique ground state is achieved as the gap between the two lowest energy levels become significant compared to the temperature in the energy scale.

\textit{Role of the next-nearest neighbor interactions.} It is believed that with the increase in the range of interactions, entanglement can be distributed among more sites in the spin chain~\cite{Daley13,Daley16,roscilde17, Ganesh21}. However, the goal of this work is to maximize entanglement between the link spins which possibly implies disentangling the \(AB:C\) bipartition from the monogamy of entanglement. Therefore, in this situation, it is not obvious to see whether the NNN interactions along with NN interactions helps in achieving the goal. We observe that in the presence of the NNN interaction, $\mathcal{P}(\lambda_m) \sim 1$ for higher $\lambda$ and, therefore, $\mathcal{E}$ achieves its maximum at a higher $\lambda_m$ although the peak value $\mathcal{E}^{\max}$ itself reduces slightly (see Fig. \ref{fig:b_2_ent_max_J2}). These effects of NNN become more pronounced at lower $\beta$ as shown in Fig. \ref{fig:b_2_ent_max_J2}. This is again due to the monogamy of entanglement since with the increase of entanglement shared among many parties, entanglement between link pairs declines. 

Let us summarize the observations in this case when $s_b=\frac{1}{2}$ and $s_\ell$ is arbitrary: (i) $\mathcal{E}^{\max}$ remains almost constant with the increase of the strength of NNN (upto a moderate $J_2$) at $\lambda_m$; (ii) $\lambda_m$ also does not change substantially with $J_2 \leq 0.2$; (iii) $\mathcal{P} (\lambda_m) \sim 1$ for $J_2 \leq 0.2$;  (iv) all the results hold for substantially low temperature, i.e., with high $\beta$. Otherwise, the $J_2$ dependence of $\mathcal{E}^{\max}$ and $\lambda_m$ is prominent for low $\beta$ values (see the \(\beta = 10^3\) case in Fig. \ref{fig:b_2_ent_max_J2}).

{\it Finite-size scaling.} It is important to determine whether the results presented above remain valid with the increase in the system size of the bulk \(C\) or not. The antiferromagnetic Heisenberg model is known to host gapless excitation in the thermodynamic limit, and hence the entanglement between two spin-$\frac{1}{2}$ sites, \(A\) and \(B\) is expected to decay with the increase in system size $N$ except at zero temperature~\cite{TasakiBook2020}. Again, we observe that although entanglement decreases with an increase in system size $N$, the decline is negligible provided the temperature is very low ($\mathcal{E}^{\max}>0.95$ for $\beta=10^4$ and $\mathcal{E}^{\max}>0.995$ for $\beta=10^5$ upto $N=16$).
The behavior of $\mathcal{E}^{\max}$ with the system size $N$ of the entire system by varying $s_\ell$ at \(A\) and \(B\) is depicted in Fig. \ref{fig:b_2_ent_max_N}.
The almost unaltered $\mathcal{E}$ for high $\beta$ can be attributed to the negligible effect of the purity $\mathcal{P}$, which shows that the requirement of extremely low temperature (close to absolute zero) for achieving two-qubit near-maximal entanglement \cite{VenutiPRA07, Dhar_2016} through a macroscopic spin-$\frac{1}{2}$ chain cannot be avoided for large system sizes by using qudits in the quantum link.

\subsection{Entangled link through bulk spin-$1$ chain}
\label{sec:spinonebulk}

Let us now move to the situation where the bulk of  spin $s_b=1$  interacts with the link spins $s_\ell$. Once again, we first assume that there is only the NN interaction among the spins $s_b=1$ in \(C\).
The scenario includes cases where \(s_b \leq s_\ell\) as well as where \(s_b > s_\ell\).  Further, we consider bilinear-biquadratic interactions both for the interspin interaction in \(C\) and also for the coupling between \(C\) and \(A\) (or \(B\)) except when $s_\ell = \frac{1}{2}$. The BBQ model of the spin-$1$ chain [Eq.~(\ref{eq:bilinbiq}) with $J_2=0$], is known to host a variety of quantum phases including Haldane phases on a periodic chain~\cite{Lai1974,sutherland1975, Babujian1982, fath1991, fath1993, bbq_prb_2022}. However, LDE properties on an open chain may not depend on the quantum phase transition points present in a periodic lattice~\cite{venuti_prl}.

\begin{figure}
    \centering
    \includegraphics[width=\linewidth]{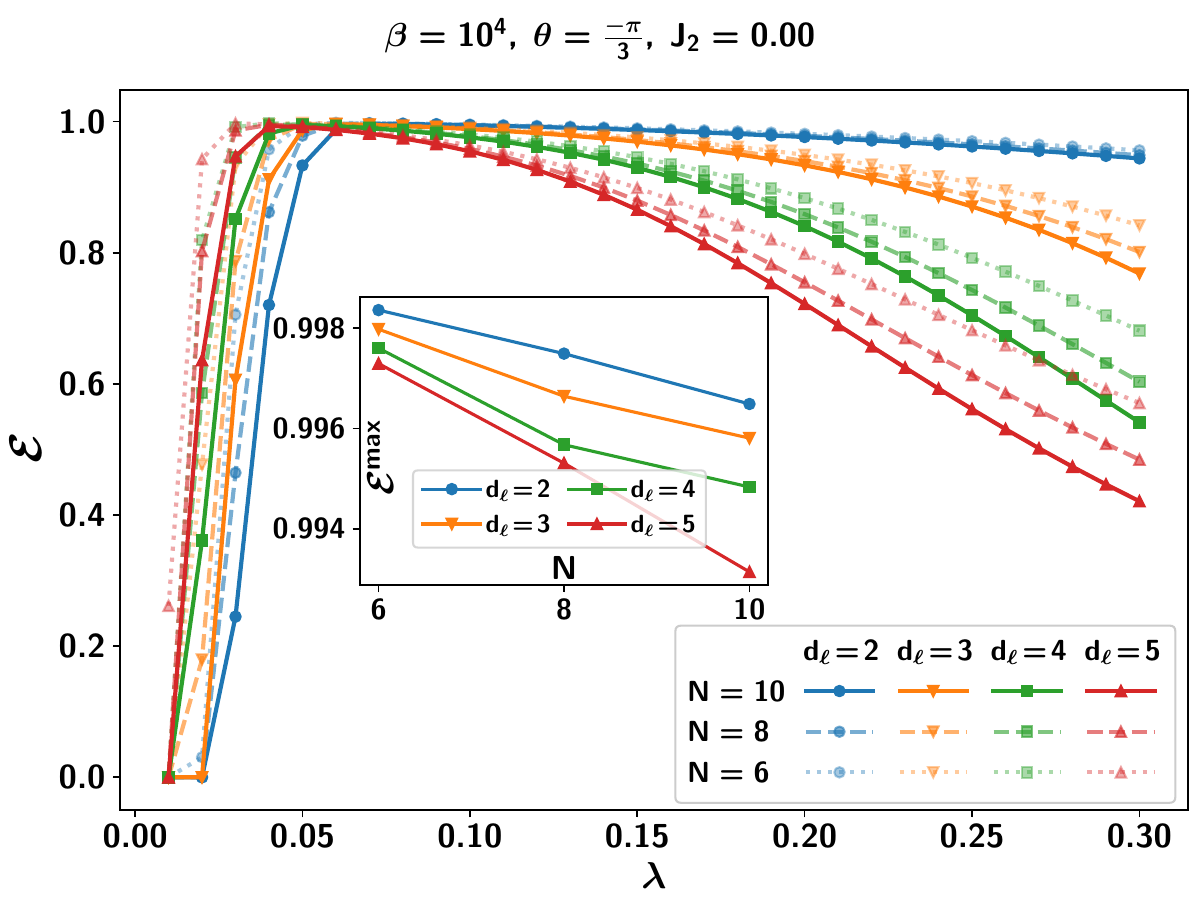}
    \caption{Two-qudit entangled link, $\mathcal{E}$ (ordinate), against coupling constant \(\lambda\) (abscissa) through a spin-$1$ bulk having only nearest-neighbor interaction (i.e., $J_2=0$). All the spins including the link sites interact according to the bilinear-biquadratic   Hamiltonian with $\theta = -\frac{\pi}{3}$. The thermal state is prepared at  $\beta = 10^4$.  Solid, dashed and dotted line represent \(N=10, 8\) and \(6\) for different individual dimension of the link  $d_\ell$ at \(A\) and \(B\). The effects of system size $N$ (abscissa) on  $\mathcal{E}^{\max}$ (ordinate) is shown in the inset. All quantities are dimensionless.}
    \label{fig:b_3_1ent}
\end{figure}

\begin{figure*}
    \centering
    \includegraphics[width=\textwidth]{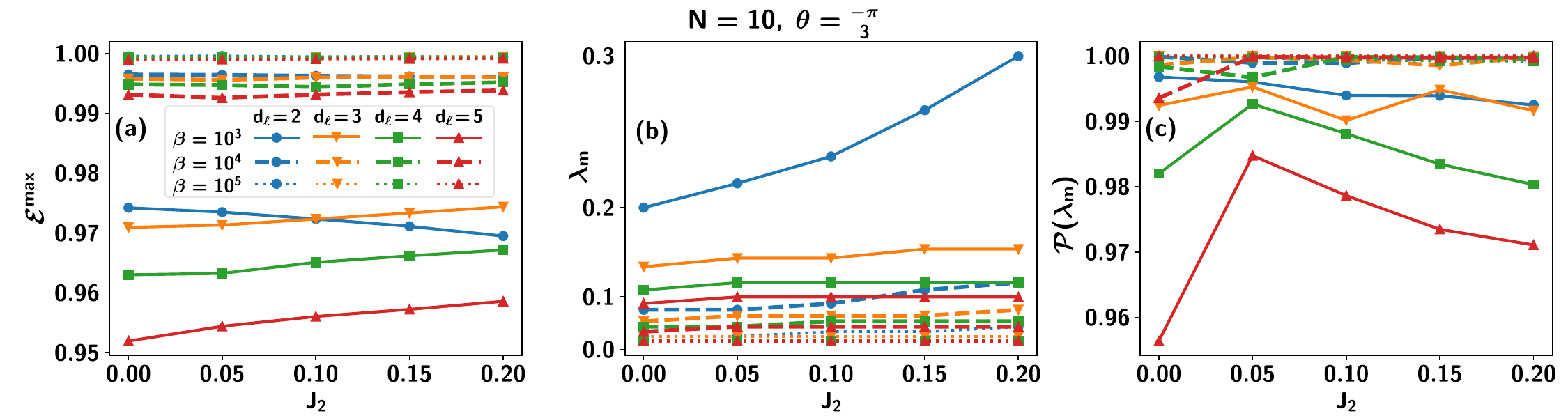}
    \caption{(a) $\mathcal{E}^{\max}$ with the variation of  the NNN interaction strength $J_2$ (abscissa) for different values of the inverse temperature $\beta$ in the spin-$1$ bulk.  (b) The link interaction strength $\lambda_m$ corresponding to $\mathcal{E}^{\max}$ (ordinate) and  (c) purity $\mathcal{P}(\lambda_m)$ (ordinate) at  $\lambda_m$. The system size is $N=10$. All the specifications in this figure is similar to the one considered for the spin-\(1/2\) bulk in Fig. \ref{fig:b_2_ent_max_J2}.  All the quantities are dimensionless. }
    \label{fig:b_3_ent_max_J2}
\end{figure*}
Here also, we study entanglement between \(A\) and \(B\) as a function of coupling strength $\lambda$ for different choices of the BBQ parameter, with $\theta$ corresponding to different quantum phases for the spin-$1$ chain (phases are known for the periodic boundary condition). We find that $\mathcal{E}^{\max}\sim 1$ for $\theta <0$ irrespective of $s_\ell$ although both $\lambda_m$ and $\mathcal{E}^{\max}$ slightly decease with $s_\ell$ (see Fig.~\ref{fig:b_3_1ent}), while for $\theta > 0$, the entanglement between \(A\) and \(B\) is very small and is highly dependent on system size even at a low temperature like $\beta = 10^4$ (see Fig.~\ref{fig:b_3_2ent} in Appendix~\ref{app:subsec:stat}). 

First, we observe that the system size dependence of $\mathcal{E}^{\max}$ is much weaker for $s_b=1, \theta=\frac{-\pi}{3}$ than $s_b=\frac{1}{2}$ ($\mathcal{E}^{\max}>0.993$ for $\beta=10^4$ till $N=10$ in Fig.~\ref{fig:b_3_1ent} inset), indicating a possibility for quasi-long-range entanglement over a longer distance with spin-$s_b(>\frac{1}{2})$ bulk. Second, for spin-$1$ bulk, the  decay of entanglement $\mathcal{E}$  in link sites with $\lambda$  is much slower than that of the spin-$\frac{1}{2}$ bulk, thereby showing the advantage of the spin-$1$ bulk state. Specifically, the survival of near-maximum entanglement, i.e., $\mathcal{E} \approx 1$  occurs for a larger range of values  of \(\lambda\) than the one observed for the spin-\(1/2\) bulk (compare Figs. \ref{fig:b_2_ent} and \ref{fig:b_3_1ent}).  It implies that the  precise control of coupling between quantum link sites and the bulk is not required in this case which clearly establishes the {\it dimensional advantage}, especially from the point of view of realization. This indicates a clear gain of using both a spin-$1$ bulk \(C\) and  bilinear-biquadratic interaction to obtain near-maximal entanglement between two distant link spins having no direct interaction. 

\textit{Temperature and the NNN dependence of entanglement.} Although in the case of spin-$\frac{1}{2}$ and spin-$1$ bulks, near-maximal entanglement can be achieved between the link spins in arbitrary dimensions, we notice that the deviation from near-maximal entanglement with the variation of temperature is substantial in the case of spin-$\frac{1}{2}$ bulk compared to the  spin-$1$ bulk (compare Fig.~\ref{fig:b_3_ent_max_J2} with Fig. \ref{fig:b_2_ent_max_J2}). However, $\mathcal{E}^{\max}$ remains unaltered with the increase of $J_2$ and so in both situations having spin $s_b (s_b=\frac{1}{2}, 1)$ bulk,  $\mathcal{E}^{\max}$, $\lambda_m$, and $\mathcal{P}(\lambda_m)$ do not change with the NNN interactions, thereby suggesting almost no role of NNN towards preparing the almost maximally entangled link (as shown in Fig.~\ref{fig:b_3_ent_max_J2}). 
More precisely, while the effect of $J_2$ is negligible for a very low temperature ($\beta \sim 10^4$ and higher), 
at a higher temperature ($\beta \sim 10^3$), small NNN interaction is found to  suppress the entanglement between \(A\) and \(B\) for $s_\ell=\frac{1}{2}$ while it mildly boosts entanglement for higher $s_\ell$. Further, NNN pushes $\lambda_m$ upward for $s_\ell=\frac{1}{2}$ while $\lambda_m$ for higher $s_\ell$ is mostly unaffected by the strength of $J_2$. In this sense, it may indicate that  long-range interaction may be beneficial for possessing LDE between two distant qudits.

\section{Proposal to create an entangled quantum link in dynamics}
\label{sec:dynamics}

To obtain a highly entangled quantum link  in the equilibrium, we show in the preceding section that extremely low temperatures are required, which are typically hard to achieve in experiments. 
Moreover, keeping the system in equilibrium is not an easy enterprise. Therefore, it is appealing to determine a situation where after evolving the system by an interacting Hamiltonian, a highly entangled quantum link can be created from the appropriate product initial state. Specifically, the bulk is prepared as a completely polarized state
by applying a high magnetic field. Therefore, the operating temperature of the protocol at which unitary dynamics can be performed efficiently can be decided independent of the preparation of the initial product state although the system has to be kept isolated during evolution. In other words, we assume that the environmental effect is negligible so that the evolution can be taken to be unitary, which has also been realized experimentally~\cite{gao_pra_2019,Imany2019,Low2020,Chi2022,Hrmo2023,liu_prx_2023,fischer_PRXQuantum_2023}.

\begin{figure}
    \centering
\includegraphics[width=\linewidth]{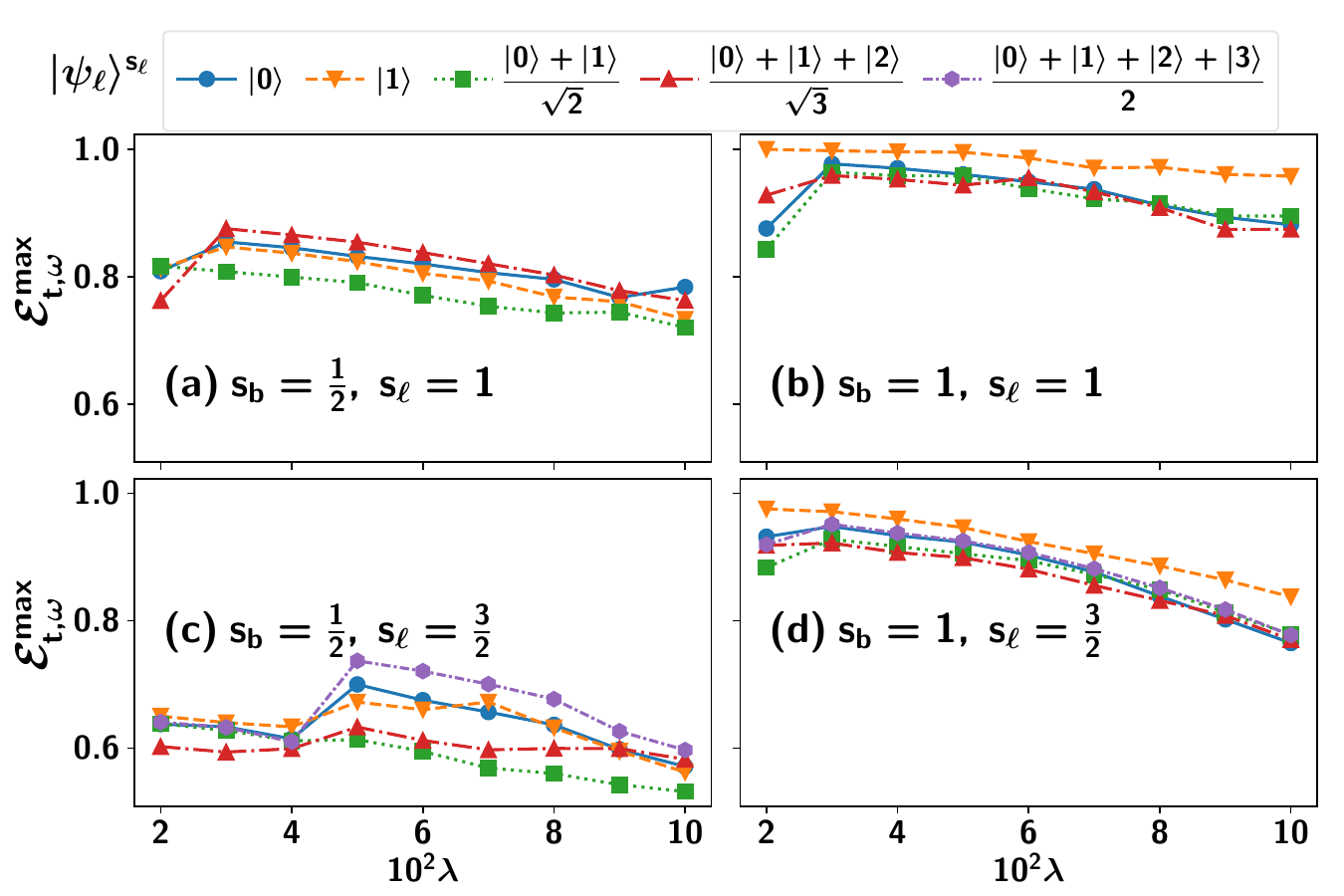}
    \caption{Role of initial state in the link. \(\mathcal{E}^{\max}_{t, \omega}\) (ordinate)  versus \(10^2 \lambda\) for different \(|\psi_\ell \rangle\) as mentioned in the legend. Different plots represent different \((s_\ell, s_b)\) pairs, with evolution by the NN Heisenberg Hamiltonian ($\theta=0, J_2=0$) $H^{(s_\ell,s_b)}(\lambda)$ of system size \(N=14\). A near-maximally entangled quantum link is obtained with \((s_\ell=1, s_b=1)\) pair and (b) initial link state $|\psi_\ell\rangle=|1\rangle$. All the quantities are dimensionless. }
    \label{fig:DG_4}
\end{figure}

{\it Choice of initial state.} Without any preferred direction of the SU(2) symmetric Hamiltonian, we choose the initial state of the bulk as $|\Psi\rangle_C=\bigotimes_{i=2}^{N-1}|0\rangle_i$ which is a full separable state.
Note that this can be prepared experimentally as a ground state of local magnetic field, i.e., $H_0=-h_0\sum_iS^z_i$. Since $H_0$ has an energy gap $E_{\text{gap}} = h_0/2$, the preparation of the pure state $|\Psi\rangle_C$ is possible even with  $\beta\sim 100$ at magnetic field $h_0\sim 1$, as $e^{-\beta E_{\text{gap}}}\sim10^{-22}$, giving a pure state for any system size of the bulk. Since such $|\Psi\rangle_C$ is an eigenstate of the SU(2) symmetric bulk Hamiltonian $H_b$ with zero entanglement (both bipartite as well as multipartite entanglement), this ensures that the entanglement generated between the link spins is purely through interactions with the bulk. The link spins $A$ and $B$ are also taken as pure states and oriented with the bulk in various directions. Hence, the initial state can be written as
\begin{equation}
    |\Psi^{\omega, \phi}(t\!=\!0)\rangle_{ABC} = |\psi_\ell^{\omega, \phi}\rangle_A\otimes \left(\bigotimes_{i=2}^{N-1}|0\rangle_i\right)\otimes |\psi_\ell^{\omega, \phi}\rangle_B,
    \label{eq:ini_st}
\end{equation}
with $|\psi_\ell^{\omega, \phi}\rangle\!\!=\!\!\hat{U}^z(\phi)\hat{U}^y(\omega) |\psi_\ell\rangle$, and $\hat{U}^\alpha(\delta)\!\!=\!\!\exp(-i\delta \hat{S}^\alpha)$ is the unitary rotation operator along the $\alpha$-axis by an angle $\delta$. $|\psi_\ell\rangle$ is the  link state
which is rotated by $\omega$ and $\phi$ describing the relative rotation of the link spins in the $y$ direction, followed by $z$ rotation with respect to the bulk, resulting in the initial link state \(|\psi_\ell^{\omega, \phi}\rangle_i\) (\(i=A, B\)). For example, taking $|\psi_\ell\rangle =|0\rangle$, the link site becomes  $|\psi_\ell^{\omega, \phi}\rangle^{s_\ell=\frac{1}{2}}=\cos\frac{\omega}{2}|0\rangle+e^{i\phi}\sin\frac{\omega}{2}|1\rangle$, while in the case of spin-$1$, it takes the form as $|\psi_\ell^{\omega, \phi}\rangle^{s_\ell=1}=e^{-i\phi}\cos^2\frac{\omega}{2}|0\rangle+\sqrt{2}\cos\frac{\omega}{2}\sin\frac{\omega}{2}|1\rangle+e^{i\phi}\sin^2\frac{\omega}{2}|2\rangle$. Note that for arbitrary spin-$s_\ell$, $|\psi^{s_\ell}(\omega,\phi)\rangle$ possesses a nontrivial form.  While for spin-$1/2$, $(\omega, \phi)$ rotations on a single state access the complete local Hilbert space, this is not true for higher spin quantum numbers.  For example, $\hat{U}^z(\phi)\hat{U}^y(\omega)|0\rangle\neq |1\rangle$, for any $(\omega, \phi)$ with $s_{\ell}>\frac{1}{2}$. Preparation of these states requires higher powers of the spin operators as local operators. Although in the case of spin-\(1/2\), we can optimize over \(\omega\) and \(\phi\) to obtain the initial link state, it is not the case for higher-dimensional systems.  Indeed, we observe that optimization over \(\omega\) and \(\phi\) provides the initial state which leads to a highly entangled quantum link, although starting from \(|1\rangle\) or \(\frac{1}{\sqrt{d_{\ell}}}\sum_{k=0}^{d_{\ell}-1} |k\rangle\) in the link also produces a highly entangled state after properly adjusting system parameters, as depicted in Fig. \ref{fig:DG_4}.  
\begin{figure}[h]
    \centering
    \includegraphics[width=\linewidth]{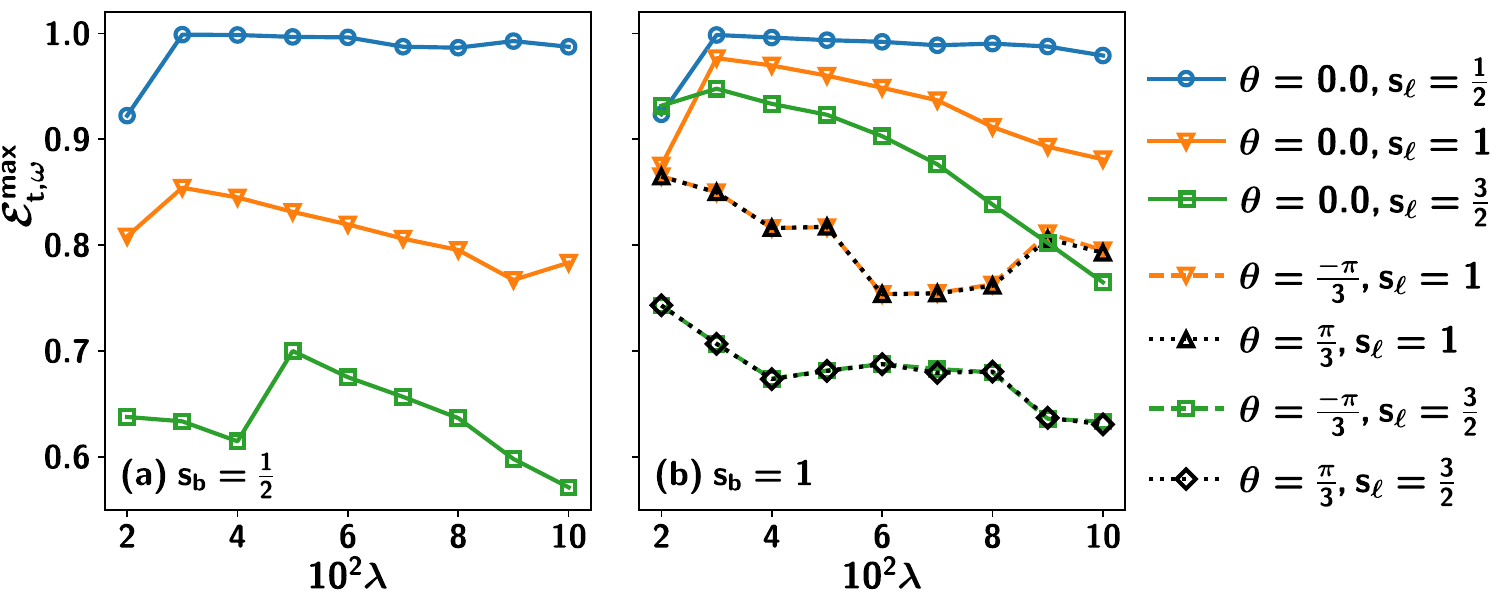}
    \caption{\(\mathcal{E}^{\max}_{t, \omega}\) (ordinate)  against \(10^2 \lambda\) (abscissa) for different values of the bulk spin quantum number, namely (a) \(s_b=1/2\) and (b) \(s_b=1/2\). The system evolves according to the bilinear-biquadratic Hamiltonian with \(\theta =0\) (dashed line), \(\theta = -\pi/3\) (solid line) and \(\theta = \pi/3\) (dotted line) from the initial state as given in Eq. (\ref{eq:ini_st}) with $|\psi_\ell\rangle=|0\rangle$ and \(N=14\). 
    It is evident that with $\theta=0$ (Heisenberg), highly entangled states between the link spins can be achieved for various values of $s_\ell$. All the quantities are dimensionless. }
    \label{fig:DG_3}
\end{figure}

{\it Dynamical states and their entanglement profile.} The initial state, prepared as in Eq. (\ref{eq:ini_st}), is evolved under the unitary evolution generated by the  SU(2) Hamiltonian as $|\Psi^{\omega, \phi}(t)\rangle_{ABC} = e^{-iH^{(s_\ell,s_b)}t}|\Psi^{\omega, \phi}(0)\rangle_{ABC}$. We observe that the entanglement in the resulting time-evolved state $|\Psi^{\omega, \phi}(t)\rangle_{ABC}$ is invariant under the rotation along the axis of the bulk, i.e., with our choice of $|\Psi\rangle_C$ only $\hat{U}^y(\omega)$ creates a nontrivial entanglement dynamics, which reduces the parameter space. Note that without any relative rotation, i.e., $\omega\!=\!0, \phi\!=\!0$, the initial state is in the highest-energy (ferromagnetically ordered) eigenstate of the evolving SU(2) symmetric antiferromagnetic model, resulting in trivial dynamics and entanglement cannot be generated between the link spins.

We find that the entanglement $\mathcal{E}$ between the link spins $A$ and $B$ oscillates with time while rotating $|\psi_\ell\rangle^{s_\ell}\!~\!=\!~\!|0\rangle$ for various values of $\omega$, with time period $\propto \lambda^{-1}$ (see Fig.~\ref{fig:dyn_3_tmp} in Appendix~\ref{app:subsec:dyn}). Therefore, we compute the maximum entanglement generated $\mathcal{E}_{t,\omega}^{\max}$ after maximizing over time to quantify the entangled link generation by $H^{(s_\ell,s_b)}(\lambda)$. We find that for small link interaction strength $\lambda\lesssim0.05$, the system size dependence of $\mathcal{E}_{t,\omega}^{\max}$ is negligible (see Fig.~\ref{fig:DG_2} in Appendix~\ref{app:subsec:dyn}).

In the case of $s_\ell>1/2$, the local Hilbert space is not completely spanned by the unitary rotations on $|\psi_\ell\rangle^{s_\ell}=|0\rangle$, for $\mathcal{E}_{t,\omega}^{\max}$. For $s_\ell=1,\) and \(3/2\), we take different starting quantum link states, $|\psi_\ell\rangle^{s_\ell}$ as shown in Fig. \ref{fig:DG_4}. Interestingly, the quantum link is highly entangled, when the link spins are initialized in the state $|\psi_\ell\rangle^{s_\ell}=|1\rangle$, with $\mathcal{E}_{t,\omega}^{\max}\sim 1$ for $s_b=s_\ell=1$, i.e., almost maximally entangled two-qutrit link can be created. This possibly indicates that the situation can be improved if one maximizes over all possible initial states of the link spins.

Let us now illustrate that the BBQ Hamiltonian can be used to dynamically create highly entangled quantum link. In particular, we compare the entanglement created between link spins with different spin-$s_b$ bulk spin and different values of the BBQ parameter $\theta$. First, with only Heisenberg interactions ($\theta=0$), we find that the maximum entanglement $\mathcal{E}_{t,\omega}^{\max}$ between the link spins decreases with increasing link spin quantum number $s_\ell$  (see Fig. \ref{fig:DG_3}) for both $s_b=1/2$ and $s_b=1$ bulk chains although the rate of decrease of $\mathcal{E}_{t,\omega}^{\max}$ with $s_\ell$ is much slower for $s_b=1$ bulk chain. In particular, for a system size $N=14$ with $s_b=1/2$,  $\mathcal{E}_{t,\omega}^{\max}\sim 1.0, 0.8, 0.6$ for $s_\ell=1/2, 1$ and $3/2$ respectively whereas for $s_b=1$, $\mathcal{E}_{t,\omega}^{\max}\sim 1.0, 0.95, 0.9$ for $s_\ell=1/2, 1$ and $3/2$ respectively (at $\lambda=0.06)$, revealing the benefit of the local spin dimension of the bulk. For a $s_b=1$ bulk chain, $\mathcal{E}_{t,\omega}^{\max}$ is lower for the BBQ parameter $\theta\neq0$, with same $\mathcal{E}_{t,\omega}^{\max}$ for $\pm\theta$. Specifically, the entanglement between the link spins reduces to $\mathcal{E}_{t,\omega}^{\max}\sim 0.8$ and $0.65$ for $\theta=\pm\pi/3$ from $0.95$ and $0.9$ for $\theta=0$ with the link spins $s_\ell=1$ and $s_\ell=3/2$ respectively. It is important to notice that in the equilibrium situation, the BBQ model with \(\theta>0\) cannot produce an entangled quantum link although it can be generated through dynamics.

\section{Conclusion}
\label{sec:conclu}

Creating a long-distance entangled quantum link can be a crucial component of quantum communication and for the scalability of a quantum computer. However, instead of two-qubit entangled states, sharing higher dimensional entangled states has been demonstrated to be beneficial for several quantum information tasks, including quantum cryptography, quantum computation and so on.

This work addresses the problem of sharing long-distance highly entangled states in arbitrary dimensions between two distant locations that are weakly coupled with a large number of interacting spins, referred to as a processor or a bulk. In an equilibrium scenario, we find that {\it a highly entangled link quantum link can be established using a spin-$\frac{1}{2}$ bulk with Heisenberg interaction as well as a spin-$1$ bulk with bilinear-biquadratic interaction.} However, the nearly maximum entanglement is found to be more robust for the spin -$1$ bulk against the variations of system parameters, including the temperature. Moreover, the presence of NNN interactions has no significant impact on the maximum entanglement for the temperature range considered in this work.

Preparing the system at a very low temperature, thereby preserving the highly entangled state over a long distance is a challenging endeavor. To overcome this, we go beyond the equilibrium situation and dynamically prepare the highly entangled quantum link. In particular, the initial state is prepared as a completely polarized fully separable state in the bulk with the quantum link states, polarized at a specific angle with respect to the orientation of the bulk spins. The system then undergoes a unitary evolution according to either the Heisenberg or the bilinear-biquadratic Hamiltonian, depending on the spin quantum number of the bulk spins. Hence, the dynamical study is performed under the assumption that the system is either not in contact with the environment or weakly interacts with the environment, which can be neglected. It results in the creation of a highly entangled quantum link.  Like the equilibrium situation, the produced highly entangled quantum link over time depends on the system size of the bulk but decreases slowly, favoring low interaction strength of link spins with the bulk for large system sizes. It will be interesting to determine the environmental conditions that allow the generation of highly entangled states to remain unaffected.

\section{Acknowledgments}
We acknowledge the support from the Interdisciplinary Cyber Physical Systems (ICPS) program of the Department of Science and Technology (DST), India, Grant No.: DST/ICPS/QuST/Theme- 1/2019/23 and TARE Grant No.: TAR/2021/000136. We acknowledge the use of \href{https://github.com/titaschanda/QIClib}{QIClib} -- a modern C++ library for general-purpose quantum information processing and quantum computing \cite{qiclib} and high-performance computing facility at the Harish-Chandra Research Institute.

\appendix
\section{$XX$ Hamiltonian with $XY$ edge interactions}
\label{app:XY}
\begin{figure*}
    \centering
    \includegraphics[width=\textwidth]{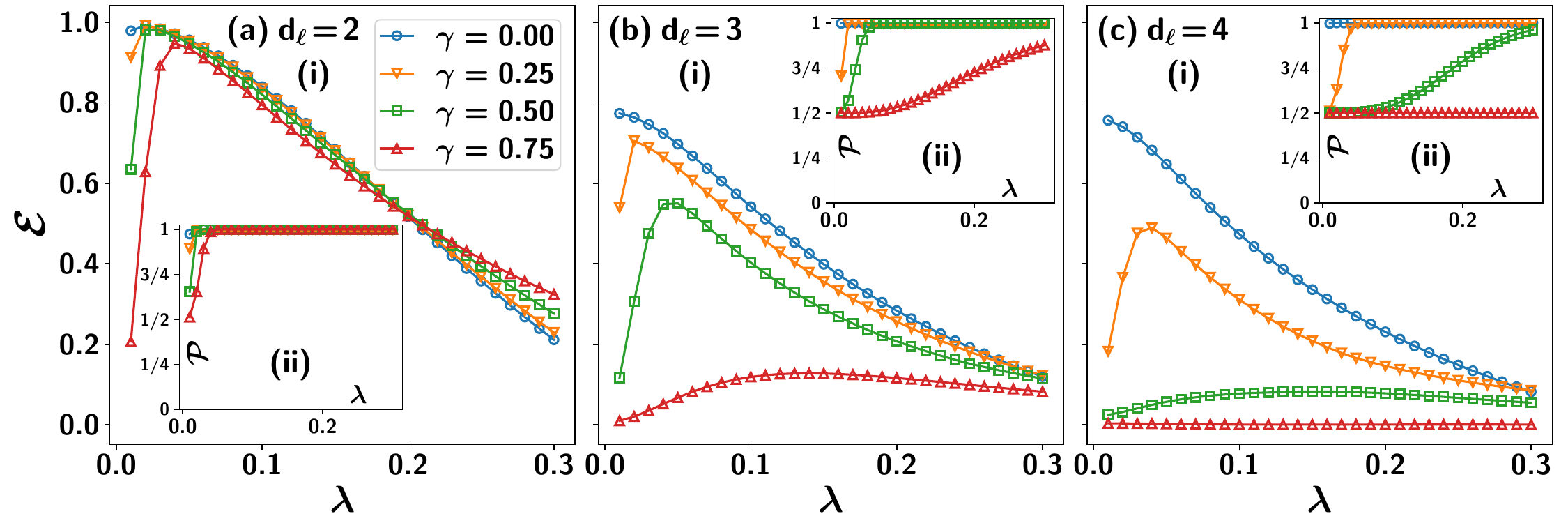}
    \caption{Entanglement $\mathcal{E}$ (ordinate) between the link spins versus link interaction strength \(\lambda\) for various link spin dimensions, namely, \(d_\ell=2, 3,\), and \(4\). $\mathcal{E}$ is generated via $XY$ interactions of link spins with the bulk, where the sites in bulk have $XX$ interactions. Inset (ii) shows the purity $\mathcal{P}$ (ordinate) against the link interaction strength $\lambda$. The system size is $N=16$, inverse temperature is $\beta=10^5$ and only the nearest-neighbor interactions are considered, i.e., $J_2=0.0$. All quantities are dimensionless.}
    \label{fig:xy_ent}
\end{figure*}

Let us consider the model with identical spin-$1/2$ bulk and link spin systems. It was shown that when the Hamiltonian of the bulk is the nearest-neighbor $XX$ chain and the link spins with $s_{\ell}=s_b=1/2$ interact with the edge spins of the bulk via $XY$ interactions, the quantum link with almost maximal entanglement can be achieved by suitably tuning the system parameters~\cite{Dhar_2016}. Specifically, $H_b=\sum_{i=2}^{N-2} \hat{S}^x_i\hat{S}^x_{i+1}+\hat{S}^y_i\hat{S}^y_{i+1}$ and $H_\ell=\lambda_1(\hat{S}^x_1\hat{S}^x_2+\hat{S}^x_{N-1}\hat{S}^x_{N}) + \lambda_2(\hat{S}^y_1\hat{S}^y_2+\hat{S}^y_{N-1}\hat{S}^y_{N})$, where $\hat{S}^\alpha_i, (\alpha=x,y)$ are spin-$1/2$ operators at site  $i$, $H_b$ describes the bulk Hamiltonian and $H_\ell$ is the Hamiltonian connecting the link to the bulk. Hence, the total Hamiltonian of the system can be represented as $H^{(\frac{1}{2},\frac{1}{2})}(\lambda, \gamma)=H^{\frac{1}{2}}_b+\lambda H^{(\frac{1}{2},\frac{1}{2})}_\ell(\gamma)$, $\left(\lambda=\frac{\lambda_1+\lambda_2}{2}, \gamma=\frac{\lambda_1-\lambda_2}{\lambda_1+\lambda_2}\right)$, with $\gamma$ as the anisotropy parameter of the connecting Hamiltonian described by the  $XY$  model. The superscript $(s_\ell=1/2,s_b=1/2)$ represents the spin quantum numbers of the link, $s_\ell$, and the bulk, $s_b$. 

Let us use the above Hamiltonian with $s_{\ell}=1$ and $3/2$ link spins. Therefore, we consider a mixed spin Hamiltonian
\begin{align*}
    H^{(s_{\ell},\frac{1}{2})}(\lambda, \gamma)\!=\!H^{\frac{1}{2}}_b\!+\!\lambda H^{(s_{\ell},\frac{1}{2})}_\ell(\gamma) \!=\!\sum_{i=2}^{N-2}\hat{S}^x_i\hat{S}^x_{i+1}\!+\!\hat{S}^y_i\hat{S}^y_{i+1} + \\ \lambda\left((1\!\!+\!\gamma)(\hat{S}^x_1\hat{S}^x_2\!+\!\hat{S}^x_{N-1}\hat{S}^x_{N}) \!+\! (1\!\!-\!\gamma)(\hat{S}^y_1\hat{S}^y_2\!+\!\hat{S}^y_{N-1}\hat{S}^y_{N})\right).
\end{align*}
Let us now keep the bulk Hamiltonian to be spin-$1/2$ systems while the link spins are taken to be spin-$s_{\ell}$. It implies that the edge spins of the bulk with which the link sites interact remain spin-$1/2$ so that the total Hamiltonian, $H^{(s_{\ell},\frac{1}{2})}_\ell$, becomes the mixed spin Hamiltonian following $XY$ interactions. We compute the thermal state of the system at inverse temperature $\beta$ (as described in Sec.~\ref{subsec:num_sim} of the main text) and compute the entanglement between the link spins $\mathcal{E}$ (defined in Sec.~\ref{subsec:ent} of the main text). It is important to note here that while for $s_{\ell}=1/2$, $\mathcal{E}\sim 1$ is achieved for various $\gamma$ at different $\lambda\ll 1$ [see Fig.~\ref{fig:xy_ent}(a)], for $s_{\ell}=1,\,\text{and}\,3/2$ we find that no tailoring of parameters $\lambda$ and $\gamma$ leads to a maximum entanglement between link spins $A$ and $B$ when $\lambda<1$ as shown in Figs.~\ref{fig:xy_ent}(b) and \ref{fig:xy_ent}(c). At $\beta=10^5$, while $\mathcal\sim 1$ is achieved for $s_{\ell}=1/2$ when the purity $\mathcal{P}\sim1$, such a behavior is not seen for higher $s_{\ell}$ (with $\mathcal{E}<1$) even when $\mathcal{P}\sim 1$. This suggests that only changing the spin-\(1/2\) link spins to a higher one may not work.

\section{Generation of entangled quantum link}

The bilinear-biquadratic Hamiltonian $H^{(s_\ell,s_b)}(\lambda)$, as defined in Eq.~(\ref{eq:bilinbiq}) of the main text, can generate significant entanglement between the qudit (spin-$s_{\ell}$) link spins mediated by a common bulk of spin-$s_b$. In this work, we explore both equilibrium and nonequilibrium scenarios: in the former, we analyze the properties of the thermal state as a function of inverse temperature $\beta$; in the latter, we consider unitary evolution  to assess the dynamic generation of entanglement. This dual approach allows us to characterize the robustness and sensitivity of entangled link formation under varying physical conditions and interaction parameters, with $s_b=1/2$ and $s_b=1$ bulk spins. We elaborate here on the scenarios in which the generation of high entanglement is affected by the system parameters.

\subsection{Long-distance two-qudit entanglement in equilibrium conditions}
\label{app:subsec:stat}

\begin{figure}[h]
    \centering
    \includegraphics[width=\linewidth]{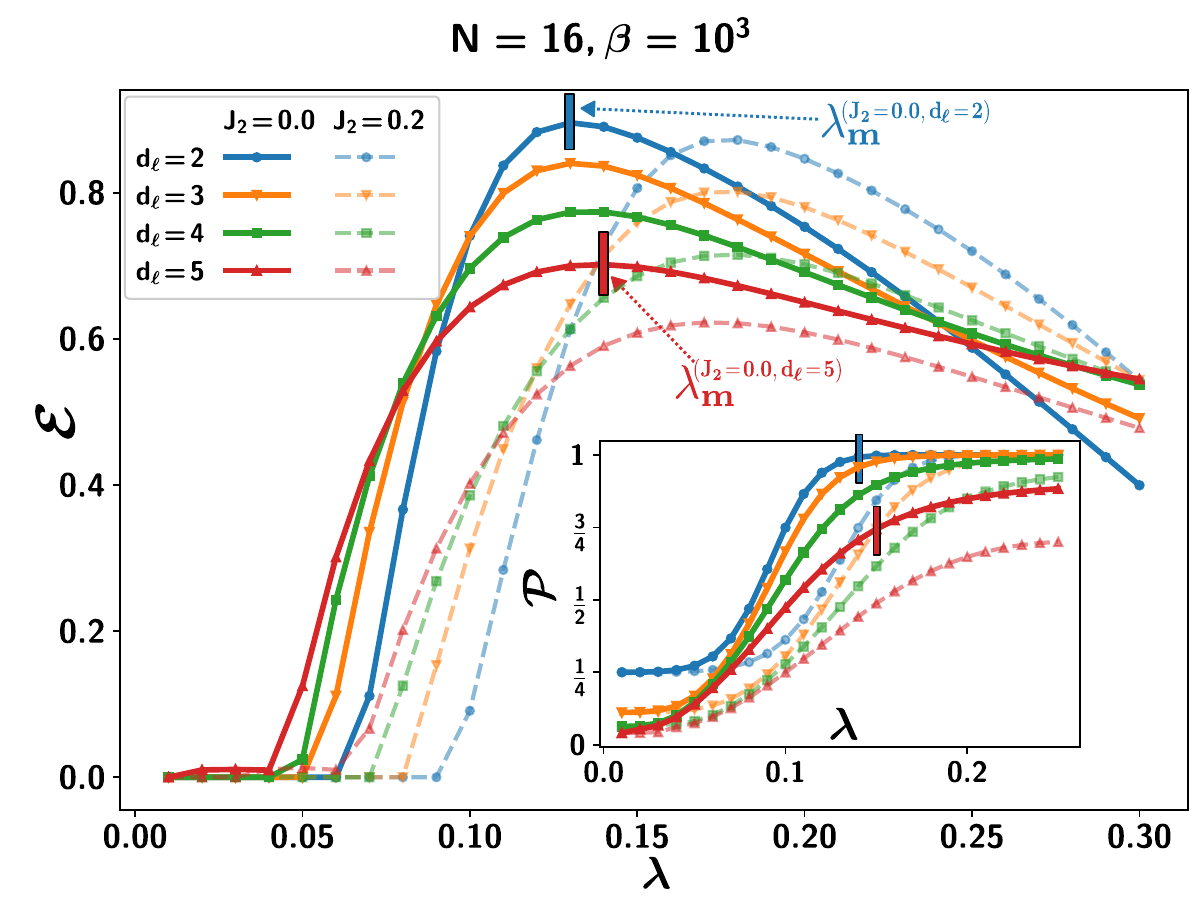}
    \caption{Entanglement $\mathcal{E}$ (ordinate) of the link versus \(\lambda\) (abscissa) for different link spin dimension $d_\ell$, with and without the next-nearest interaction strength $J_2$. All other specifications are same as in Fig. \ref{fig:b_2_ent} of the main text.
    (Inset) Purity $\mathcal{P}$ for the corresponding system with \(\lambda\) having vertical bars at $\lambda_m$. Here $N=16, \beta=10^3$, and all quantities are dimensionless.}
    \label{fig:b_2_1_ent}
\end{figure}

Let us first discuss the effects of low inverse temperature $\beta$ on the entanglement $\mathcal{E}$  between the link spins with $s_b=1/2$ bulk for arbitrary \(s_{\ell}\). While with $\beta=10^4$, $\mathcal{E}^{\max}\sim 0.99$ for all spin-$s_{\ell}$ link spins considered (Fig.~\ref{fig:b_2_ent} of the main text), the situation is degraded on increasing the temperature of the system, i.e., lowering $\beta$. Specifically, for $\beta=10^3$, $\mathcal{E}^{\max}\sim 0.87$ for spin-$1/2$ link spins, and decreases substantially to $\mathcal{E}^{\max}\sim 0.6$ for spin-$2$ link spins as shown in Fig.~\ref{fig:b_2_1_ent}. This can be seen as a result of decrease in purity $\mathcal{P}$ of the state for small link interaction strength $\lambda$, and  $\mathcal{E}^{\max}$ is obtained at higher $\lambda$, when $\mathcal{P}\sim 1$. The ground state at higher $\lambda$ possess low $\mathcal{E}$, decreasing $\mathcal{E}^{\max}$ obtained between the link spins $A$ and $B$.

\begin{figure}[h]
    \centering
    \includegraphics[width=\linewidth]{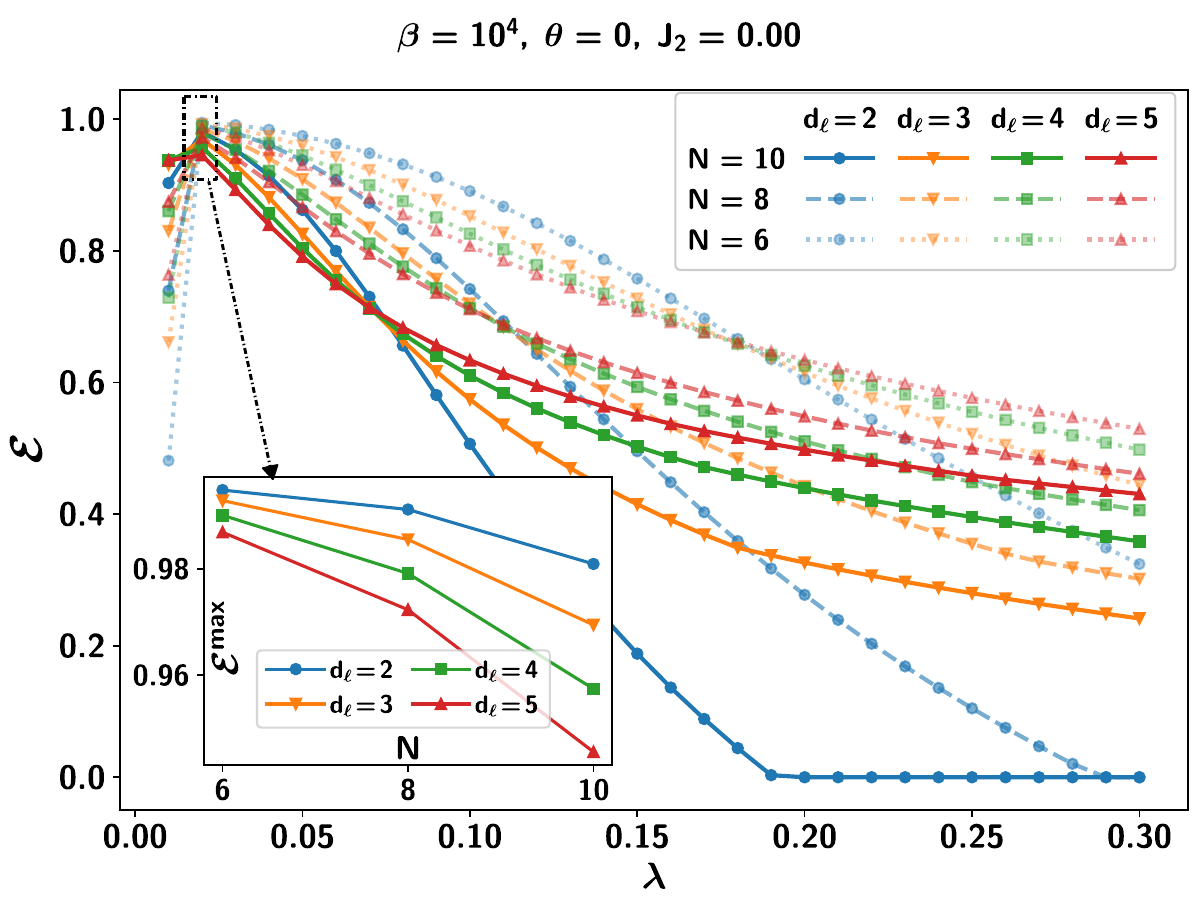}
    \caption{$\mathcal{E}$ (ordinate) as a function of the coupling strength $\lambda$ (abscissa) for the spin-\(1\) bulk. The interacting Hamiltonian in this case is chosen to be the Heisenberg interaction, i.e., $\theta = 0$ at $\beta = 10^4$. (Inset) Maximum entanglement $\mathcal{E}^{\max}$ (ordinate) with system size $N$ (abscissa). All quantities are dimensionless.}
    \label{fig:b_3_2ent}
\end{figure}

In case of the system with spin-$1$ bulk ($s_b=1$), the BBQ parameter $\theta$ can influence the entanglement between the link spins via $H^{(s_\ell,s_b)}(\lambda)$. Specifically, for Heisenberg-type interspin interaction, i.e., $\theta=0$, the maximum entanglement between the link spins $A$ and $B$, $\mathcal{E}^{\max}>0.95$ is obtained for all $s_{\ell}\leq2$, at small interaction strengths only ($\lambda_m\sim0.02$). Note further that $\mathcal{E}^{\max}$ exhibits stronger dependence on \(\lambda\) for Heisenberg-type interspin interaction (see Fig.~\ref{fig:b_3_2ent} for $\theta=0$) than the $\theta<0$ case since for $\theta=0$, $\mathcal{E}$ again decreases faster from its maximum as $\lambda$ is increased from $\lambda_m$ compared to the case with \(\theta <0\) (comparing with Fig. \ref{fig:b_3_1ent} of main text). Therefore, with $\theta<0$, bilinear-biquadratic interactions can create a more stable  highly entangled quantum link in terms of tuning of \(\lambda\) than that obtained via Heisenberg model.

\subsection{Entanglement generation in dynamics}
\label{app:subsec:dyn}

Let us illustrate the dynamical protocol, given in Sec.~\ref{sec:dynamics} of the main text, with initial link states, $|\psi_\ell\rangle^{s_\ell}\!~\!=\!~\!|0\rangle$ for various $(s_\ell, s_b)$ pairs. Hence, the initial state of the entire system becomes \(=\!~\!|0\rangle_A \otimes_{i=2}^{N-1} |0\rangle_i \otimes |0\rangle_B \) which evolves according to the Hamiltonian \(H^{(s_\ell, s_b)}(\lambda)\). After tracing out the bulk \(C\) from the dynamical state $|\Psi^{\omega, \phi}(t)\rangle_{ABC}$, the quantum link spins $A$ and $B$ are entangled with entanglement $\mathcal{E}$ increasing initially with time $t$.


Before presenting the results, let us elaborate on the difference between the quantum state transfer protocol~\cite{Bose} and the dynamical protocol considered here. 
In the quantum state transfer protocol~\cite{Wojcik05, Bayat2011}, the initial state is an $(N+1)$-party state, $\ket{\Psi_{\text{qst}}(0)}_{N+1}=|\Phi^+\rangle_{1^\prime1}\otimes \left(\bigotimes_{k=2}^{N-1}|0\rangle_k\right)\otimes |0\rangle_N$, where $|\Phi^+\rangle_{1^\prime1}$ can be a Bell state ({\it maximally entangled state}) between sites $1^\prime$ and $1$, and the sites from $k=1$ to $N$ undergoes unitary dynamics with $N$-site Hamiltonian $H$, while the $k=1^\prime$ site does not go under any evolution. Therefore, the overall evolution is given by $\ket{\Psi_{\text{qst}}(t)}_{N+1}=\left(\mathbb{I}_{1^\prime}\otimes e^{-iHt}\right)\ket{\Psi_{\text{qst}}(0)}_{N+1}$. The quantum state transfer is considered successful when the sites $1^\prime$ and $N$ in $\ket{\Psi_{\text{qst}}(t)}_{N+1}$ are (near) maximally entangled, exhibiting transfer of quantum information (and correlations) from site $1$ to $N$. Note that the quantum state is always maximally entangled between the $\{1^\prime\}:\{1,2,\dots,N\}$ bipartition, and the sites $1$ and $N$ in $\ket{\Psi_{\text{qst}}(t)}_{N+1}$ are not entangled. Therefore, an already existing entanglement is {\it transported} in quantum state transfer.

However,  we consider an initial {\it product state} of $N$ sites, $\ket{\Psi(0)}_{N}=|\psi\rangle_{1}\otimes \left(\bigotimes_{k=2}^{N-1}|0\rangle_k\right)\otimes |\psi\rangle_N$, and the entire system is evolves unitarily via the $N$-site Hamiltonian $H$, with the evolution given as $\ket{\Psi(t)}_{N}=e^{-iHt}\ket{\Psi(0)}_{N}$. Our aim is to generate (near) maximal entanglement between the sites $1$ and $N$, which can also be of different spin-quantum-number than the bulk (sites $\{2,3\dots,N-1\}$. Note that the initial state does not possess any correlation, and the interaction between \(1\) and \(2\) as well as \(N-1\) and \(N\) are considered to be different from the interactions between \(2\) and \(N-1\) (which we refer to as bulk). Therefore, we study the {\it generation} of entanglement between distant sites through interactions with the bulk with the help of weak interactions between sites to be entangled and the bulk.

\begin{figure}[h]
    \centering
    \includegraphics[width=\linewidth]{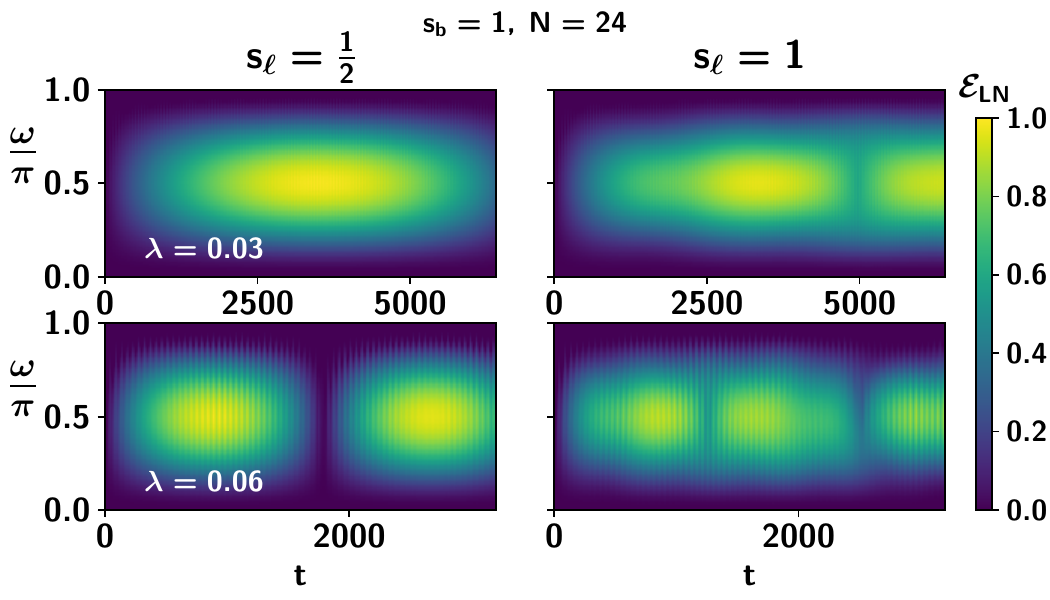}
    \caption{Map plots of entanglement \(\mathcal{E}\) between the link spins with the variation of \(t\) (horizontal axis) and \(\frac{\omega}{\pi}\) (vertical axis). Note that \(\omega/\pi\) rotation helps to create the suitable initial state in the link. The upper panels are for \(\lambda =0.03\) while the lower panels are for \(\lambda =0.06\). The initial state of the bulk is taken to be \(|0\rangle^{\otimes N-2}\) and \(s_b=1\), with the link spins taken as rotation of $|\psi_\ell\rangle^{s_\ell}=|0\rangle$. The evolution is performed with the NN Heisenberg Hamiltonian ($\theta=0, J_2=0$) for system size $N=24$. In the left column, \(s_\ell =1/2\) and in the right one, \(s_\ell =1\). All the quantities are dimensionless.  }
    \label{fig:dyn_3_tmp}
\end{figure}

Figure~\ref{fig:dyn_3_tmp} illustrates the entanglement dynamics when the evolution is generated by the nearest-neighbor Heisenberg Hamiltonian ($\theta=0, J_2=0$) with the spin-$1$ bulk. The entanglement $\mathcal{E}$ between the link spins $A$ and $B$ oscillates with time, and a highly entangled quantum link during a period of time can be produced for various spin dimensions of the link site. After oscillations, the quantum link is disentangled again in the timescale $\propto \lambda^{-1}$ followed by increased entanglement (see Fig.~\ref{fig:dyn_3_tmp}). We also observe that the maximum amount of entanglement is generated between the link spins, when $\omega\sim \pi/2$. The timescale of oscillations is observed to be increasing with system size $N$ which leads to a finite-size study.

\begin{figure}[h]
    \vspace{5mm}
    \centering
        \includegraphics[width=\linewidth]{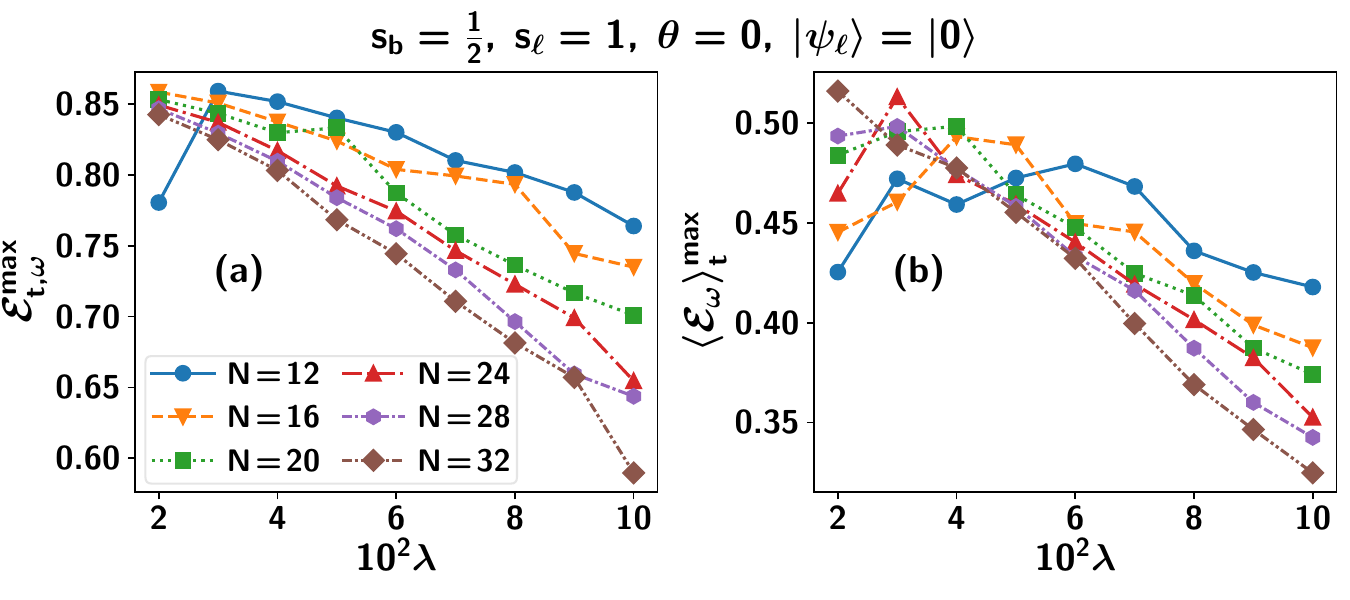}
    \caption{(a) \(\mathcal{E}^{\max}_{t, \omega}\) (ordinate) and (b) \(\langle \mathcal{E}_{\omega}\rangle_t^{\max}\) against \(10^2 \lambda\) (abscissa) for different values of the system size. The bulk is chosen to be \(|0\rangle ^{\otimes N-2}\). Here \(s_b =1/2\) and the evolving Hamiltonian is the NN Heisenberg model ($\theta=0, J_2=0$).
    All the quantities are dimensionless. }
    \label{fig:DG_2}
\end{figure}

{\it Scaling analysis.} We compute the maximum entanglement generated $\mathcal{E}_{t,\omega}^{\max}$ and average entanglement generated over time $\langle\mathcal{E}_{\omega}\rangle^{\max}_t$. \(\mathcal{E}^{\max}_{t, \omega}\) is obtained after maximizing over time, \(t\) and over \(\omega\) involved in the rotation provided the initial state of the link is chosen to be \(|\psi_\ell \rangle = |0\rangle\). In the case of computing \(\langle \mathcal{E}_{\omega}\rangle_t^{\max}\), we perform maximization over \(\omega\) after the averaging is performed over time, \(t\). Interestingly, while both the quantities decrease with system size, the dependence is small for low interaction strength ($\lambda\lesssim0.05$), shown for $|\psi_\ell\rangle=|0\rangle$ with $s_b=1/2$ and $s_\ell=1$ in Fig. \ref{fig:DG_2}. For this ($s_\ell$, $s_b$) pair, $\mathcal{E}_{t,\omega}^{\max}>0.75$ upto $N=32$, with $\lambda\leq0.04$ for the spin-$1$ link spins. In the main text, we show that the higher entangled link $\mathcal{E}^{\max}\gtrsim0.95$ can be achieved by choosing different initial link spin states $|\psi_\ell\rangle^{s_\ell}$ for $s_{\ell}\geq1$.

\bibliography{ref.bib}

\end{document}